\documentclass[prc,aps,amsfonts,twocolumn,dvips]{revtex4}
\usepackage{graphics,epsfig,amssymb}

\begin{document}

\title{\bf
Cooper pair sizes in superfluid nuclei in a simplified model}

\author{\rm X. Vi\~nas$^{(a)}$, P. Schuck$^{(b,c,d)}$, 
N. Pillet$^{(e)}$,
%J.-F. Berger$^{(e)}$, 
%N. Sandulescu$^{(b,e,f)}$} 
}

\bigskip

\address{\rm
$^{(a)}$~ Departament d'Estructura i Constituents
de la Mat\`eria and Institut de Ci\`encies del Cosmos,
Facultat de F\'{\i}sica, Universitat de Barcelona,
Diagonal {\sl 647}, {\sl 08028} Barcelona, Spain\\
$^{(b)}$~  Institut de Physique Nucl\'eaire, CNRS, 
UMR8608, Orsay,
F-91406, France \\
$^{(c)}$~  Universit\'e Paris-Sud, Orsay,
F-91505, France\\
$^{(d)}$~ Laboratoire de Physique et Mod\'elisation des 
Milieux Condens\'es,
CNRS and Universit\'e Joseph Fourier, Maison des Magist\`eres, 
Bo\^{\i}te Postale 166, 38042 Grenoble Cedex, France \\
$^{(e)}$~ CEA/DAM/DIF, F-91297 Arpajon, France \\
%$^{(e)}$~ DPTA/Service de Physique nucl\'eaire, CEA/DAM 
%Ile de France, BP12, F-91680 Bruy\`eres-le-Ch\^atel, France \\
%$^{(f)}$ Institute of Physics and Nuclear Engineering, 
%76900 Bucharest,Romania 
}

\def\fid{\vert\phi >}
\def\fig{< \phi\vert}
\def\psid{\vert\Psi>}
\def\psig{<\Psi\vert}
\def\psid{\vert\Psi>}
\def\psig{<\Psi\vert}
\def\dspt{\displaystyle}

\begin{abstract}
Cooper pair sizes are evaluated in a simple harmonic oscillator model 
reproducing the values of sophisticated HFB calculations. Underlying reasons 
for the very small sizes of 2.0-2.5 fm of Cooper pairs in the 
surface of nuclei 
are analysed. It is shown that the confining properties of the nuclear volume 
is the dominating effect. It is argued that for Cooper pair sizes 
the LDA idea is particularly inadapted.
\end{abstract}

\maketitle

Recent studies have revealed surprisingly small extensions of Cooper pairs in 
the surface of superfluid nuclei \cite{pil07,pas08,mat05}.
Such features are potentially very 
important in pair transfers in nuclear reactions \cite{oer01}.
Though the reason for the 
small sizes has been identified in our preceding paper \cite{pil10} 
to be due to the finite size of 
nuclei, it is nevertheless instructive to further elaborate on the underlying 
reasons of this behavior. We, therefore, develope a simplified model 
which, however, will keep all the essential ingredients for the comprehension 
of the effect. The model consists of a spherical harmonic oscillator (HO) 
potential (without spin-orbit) for the mean field  together with a realistic 
treatment of pairing using the Gogny D1S force \cite{D1S}. 
We will see that such a model quite accurately reproduces the 
results for the so-called coherence length (CL), i.e. the size of Cooper 
pairs, of much more sophisticated 
selfconsistent HFB calculations \cite{pil07}.

The questions we will try to answer are the following:

i) What is the reason for the existence of such very small sized Cooper pairs 
with extensions 2.0-2.5 fm in the surface of nuclei, about a 
factor 2-3 times smaller than the smallest size in nuclear matter at 
low densities? 
Those values are also much smaller than the ones estimated from the 
common believe that Cooper pair 
sizes in nuclei are of about the nucleus' dimension \cite{BM75}, 
what is based on  pairing in nuclei being in 
the weak coupling regime. Since in weak coupling CL$>b$, with $b$ the 
oscillator length to be used below, the fact that for a nucleus with, e.g. 
nucleon number $A$=120, $b \sim$ 2.2 fm $\sim $ CL$_{min.}$, does it lead to the 
conclusion that in the surface pairing is close to strong coupling? Those 
small sizes also  
are of similar magnitude as the one 
of the deuteron, that is a bound state. Does it mean that the neutron Cooper 
pairs are locally also eventually in a bound state? Actually, this might 
not be completely surprising, since two neutrons are almost bound even in 
free space and pairing could help to make them truely bound.
The question then is whether the 
small size of the CL's is due to particularly strong pairing in the 
nuclear surface 
(local strong coupling) or whether it is
essentially due to the confining constraints from the nuclear volume? It 
will be shown that the small sizes are dominantly due to the latter effect.

ii) The minimum of the CL, $\xi(R)$, in LDA is about at the same 
density as the one 
in the quantal case \cite{pil10}. Then, is the qualitative 
resemblence of $\xi(R)$ calculated 
from nuclear matter 
in LDA and the quantal $\xi(R)$ a fortuitous coincidence, or is that a 
manifestation of similar pairing correlations in both cases? We will see 
that the quantal behavior of $\xi(R)$ in finite nuclei is very similar 
for nominal and almost vanishing 
pairing. In the latter case one should not talk about coherence length but 
simply of the rms distance of uncorrelated pairs coupled to angular 
momentum L=0 
which is entirely determined by the single particle mean field 
wave functions.

We begin our considerations with the density matrix corresponding to 
one major shell of a spherical HO potential  
$V(R) = \frac{m}{2}\omega^2R^2$ with $\hbar \omega = 41A^{-1/3}$ MeV 

\begin{equation}
\hat \rho_N = {\sum_{nlm}}^{'} |nlm><nlm|,
\end{equation}

\noindent
where the prime on the sum indicates that it only runs over 
all the states $|nlm >$ contained in the major shell N.

We start out transforming this density matrix into Wigner (W) space. W-space, 
or phase space, is useful for certain aspects and furthermore it has a well 
known analytic form for the case of a HO potential where it only depends on 
the classical Hamiltonian $H_{cl.} = p^2/2m + V(R)$. The corresponding 
W-distribution is given by \cite{pra81}

\begin{equation}
\hat \rho_N|_{W} = f_N(H_{cl.}) = 8 (-1)^N e^{-\frac{2H_{cl.}}{\hbar 
\omega}} L_N^{(2)}\bigg(\frac{4H_{cl.}}{\hbar \omega}\bigg),
\label{eq2} \end{equation}

\noindent
where $L^{(\lambda)}_n(x)$ are the generalized Laguerre polynomials.
%The Wigner distribution of the total density matrix is then obtained by 
%\cite{pra81}

%\begin{equation}
%f(H_{cl.}) = \sum_{N=0}^{N=N_F} f_N(H_{cl.}),
%\end{equation}

%\noindent
%where $N_F$ corresponds to the major shell at the Fermi energy.
%This phase space distribution is shown as a function 
%of $H_{cl.}$ in Fig. 1 for the case 
%of a hypothetical symmetric nucleus with $2N=2Z=A=120$. 
%In addition, we show in Fig. 1   
%the Thomas Fermi (TF) distribution.
%\cite{pra81}. Furthermore, 
%we also present in Fig. 2 $\vert f_{N_F}(H_{cl.})\vert^2$ and its derivative 
%$\vert df_{N_F}(H_{cl.})/dH_{cl.}\vert^2$, to be used below.

%\begin{figure}
%\includegraphics[width=7.5cm,angle=-90]{FHO120g.ps}
%\caption{(Color online) Wigner distribution function for a symmetric 
%nucleus with A=120 nucleons computed quantally (solid line) and 
%obtained in the Thomas-Fermi approximation (dashed-dotted 
%line)}
%\label{Figure1} 
%\end{figure}

%\begin{figure}
%\includegraphics[width=7.5cm,angle=-90]{KappaRS5new.ps}
%\caption{(Color online) Square of the modulus of the on-shell density 
%($|f_{N_F}(H_{cl.})|^2$) and 
%of its derivative respect to $H_{cl.}$ ($|df_{N_F}(H_{cl.}/dH_{cl.}|^2$) 
%correponding to the major shell at the Fermi energy $N_F$=3 for a symmetric 
%nucleus of A=120 
%nucleons.} 
%\label{Figure2}
%\end{figure}

We are now ready to present our simplified pairing model. We shall write 
the W-transform \cite{RS80} 
of the anomalous density matrix 
$\kappa({\bf r},{\bf r}') = \\  <BCS|a^+({\bf r})a^+({\bf r}')|BCS>$ as 
( spin singlet wave function is suppressed)

\begin{equation}
\kappa({\bf R},{\bf p}) = \sum_N \kappa_N f_{N}(H_{cl.}),
\label{eq4}
\end{equation}

\noindent
with $\kappa_N = u_Nv_N$ and $u_N,v_N$ the usual BCS amplitudes. Please 
note 
that the degeneracy factors are missing in Eq. (\ref{eq4}). This stems 
from the fact 
that expression (\ref{eq2}) is not normalised to unity but to the 
degeneracy of the shell $N$.

\noindent
The gap parameters $\Delta_N$ can be obtained from the solution of 
a gap equation with matrix elements averaged over major shells 
\cite{vin03}.

\begin{equation}
\Delta_N = \sum_{N'} D_{N'}V_{N,N'} 
\frac{\Delta_{N'}}{2\sqrt{(E_{N'} - \mu)^2 + \Delta^2_{N'}}},
\end{equation}

\noindent
where $E_N = (3/2 + N) \hbar \omega$,  $D_N = 
(N+1)(N+2)/2$ is the degeneracy factor of major shell $N$, and $V_{N,N'}$ is 
the shell-averaged pairing matrix element 
To obtain $V_{N,N}$ we start from the state-dependent 
pairing matrix element \cite{RS80}
\begin{eqnarray}
&&<\Phi(\nu,\bar{\nu})|v|\Phi(\nu',\bar{\nu}')> =
\nonumber \\
&&<\nu,\bar{\nu}|v|\nu',\bar{\nu}'> - 
<\nu,\bar{\nu}|v|\bar{\nu}',\nu'>,
\label{pmatel}
\end{eqnarray}
where the two-particle states $|\nu,\bar{\nu}>$ are product states
$|\nu>$ and $|\bar{\nu}>$. The states $|\nu>$ are represented by 
single-particle wave-functions $\phi_{\nu}(\vec{r},\sigma)=
\phi_{nlm}(\vec{r}) \psi_{\sigma}$ and the corresponding time reversal 
states by
$\phi_{\bar{\nu}}(\vec{r'},\sigma)=(-1)^{1/2-\sigma} 
\phi^*_{nlm}(\vec{r'})\psi_{-\sigma}$.
Averaging over the energy  shells $E_N$ and $E_N'$ , it is easy to show 
that in phase space the shell-averaged pairing matrix elements read 
(see \cite{vin03} for more details.). 

\begin{eqnarray}
V_{N,N'} &=& \frac{1}{D_N D_{N'}}\int d^3R \int \frac{d^3 p 
d^3 p'}{(2\pi \hbar)^6} f_N(H_{cl.}) 
f_{N'}(H'_{cl.})\nonumber \\ 
&& \times v_{\eta}({\bf p} - {\bf p'}),
\end{eqnarray}

\noindent
with $v_{\eta}(p)=\eta v(p)$ and $v(p)$ being the Fourier 
transform of the Gogny 
D1S interaction in the $^1$S$_0$ pairing channel \cite{D1S}.  The 
factor $\eta$ serves 
to adjust the pairing intensity by 
hand. 

%On average such matrix elements very well reproduce 
%the quantal values [..]. We then can solve the gap eq (6) 
%to obtain $\Delta (E)$. 
%It also can roughly be fitted by a Fermi function 
%[..]
%
%
%\begin{equation}
%\Delta(E) = \frac{\Delta_0}{1+e^{(E-E_F)/a}}
%\end{equation}

%\noindent
%with $\Delta_0 = ..., E_F = ...$, and $a = ...$. Since we want to study the 
%coherence length or extension of the Cooper pairs as a function of the 
%pairing strength, we then simply write

%\begin{equation}
%\kappa_{\alpha}(E) = \frac{\Delta_{\alpha}(E)}{2\sqrt{(E-\mu)^2 + 
%\Delta_{\alpha}^2(E)}}
%\end{equation}

%\noindent
%with $\Delta_{\alpha}(E) = \alpha \Delta(E)$ and $\alpha$ a scaling factor 
%ranging between $0< \alpha < 1$. In this context, it should be realised that 
%for $\Delta_{\alpha}/\mu << 1$, $\kappa(E)$ is very much concentrated 
%around $E=\mu$ with width $\sim \Delta_{\alpha}$.\\

%In the $\gamma=0$ limit above equations can be written as

%\begin{equation}
%\Delta_N = \sum_{N'} g(E_{N'})V_{N,N'} 
%\frac{\Delta_{N'}}{2\sqrt{(E_{N'} - \mu)^2 + \Delta^2_{N'}}}
%\end{equation}

\begin{figure}
\includegraphics[width=7.5cm,angle=-90]{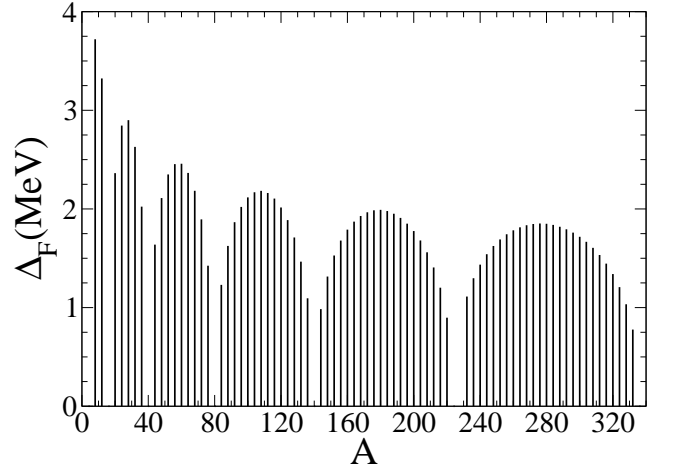} 
\caption{Pairing gap at the Fermi energy computed using the 
Gogny D1S force as a function of the number of nucleons A in an isotropic 
HO potential.}
\label{Figure1}
\end{figure}

\noindent
In Fig. \ref{Figure1}, we 
give the gap at the Fermi energy $\Delta_{F}$ as a function of A. 
We take $\eta = 0.85$ to compensate for the fact that  
we use the bare mass, 
$m^*=m$ what usually overestimates pairing. We see that the typical arch 
structure 
is recovered. Without any averaging, the gap values would depend on the 
individual single particle quantum numbers $n, l$ and Fig. 1 would show an 
additional fine structure. In the present case an averaging over the 
individual substates of one major shell has been performed not wyping out, 
however, the essential  
quantum features. 

%Let us also mention that the use of the Strutinsky averaged distribution 
%shown in Fig. 1 in the context of pairing calculations, turns our approach 
%into a 
%continuum model changing discrete 
%sums in eqs (4-5) into integrals  \cite{vinas-farine}. For instance Eq. 
%(4) 
%takes the following form

%\begin{equation}
%\kappa({\bf R}, {\bf p})= \int_0^{\infty} dE \kappa(E) \tilde f_E(H_{cl.})
%\end{equation}

%\noindent
%where $\tilde f_E$ is the Strutinsky averaged on energy shell distribution 
%function 
%and $\kappa(E)$ is the continuum version of the pairing tensor \cite{g}. 
%It also is known that the Strutinsky method is very close to the semiclassical 
%approach of Wigner and Kirkwood \cite{RS80}. How pairing and in particular 
%the CL can accurately be treated semiclassically will be published elsewhere 
%\cite{vinas-farine}.
%We will not 
%give any further details here of the Strutinsky method, since it is well 
%documented in the literature \cite{RS80} and we only will need it for 
%qualitative argumentation.

We now proceed to the calculation of the CL.
Interpreting the anomalous density as the wave function of a Cooper pair 
( we are aware of the fact that this point of view has been debated 
recently \cite{duke}), 
the local rms value of a pair is given by \cite{pil07}

\begin{eqnarray}
\xi(R)&&\equiv \sqrt{\frac{N(R)}{D(R)}} 
= \sqrt {\frac{\int d^3s s^2 
\kappa^2({\bf R},{\bf s})}
{\int d^3s \kappa^2(({\bf R}, {\bf s})}}\nonumber \\
&=& \sqrt{ \frac{ \int \frac{d^3p}{(2\pi \hbar)^3} 
|d\kappa(H_{cl.})/d(p/\hbar)|^2}
{\int \frac{d^3p}{(2\pi \hbar)^3} \kappa^2(H_{cl.})}}.
\label{CL}
\end{eqnarray}
\noindent
Here $2{\bf R} = {\bf r} + {\bf r}'$ and ${\bf s} = {\bf r} - {\bf r}'$ 
and $\kappa({\bf R}, {\bf s})$ is the Fourier transform 
of $\kappa({\bf R}, {\bf p})$ of (3).

Using Eqs. (\ref{eq2}) and (\ref{eq4}), denominator and numerator under 
the square root in 
Eq.(\ref{CL}) can be 
obtained explicitly in the 
case of the HO potential:
\begin{eqnarray}
&&D(R) = \frac{4 \alpha^3}{\pi^2}\sqrt{\frac{\pi}{2}} e^{- 2 \alpha^2 R^2}
\sum_K \sum_J (-1)^{K+J} \kappa_K \kappa_J \nonumber \\ 
&& \hspace{-0.6cm} \times \sum_{K1=0}^{min(K,J)}
L^{(1/2)}_{K1}(0) L^{(1/2)}_{K-K1}(2 \alpha^2 R^2)
L^{(1/2)}_{J-K1}(2 \alpha^2 R^2),
%\nonumber \\
\end{eqnarray}
\begin{eqnarray}
&&N(R) = \frac{12 \alpha}{\pi^2} \sqrt{\frac{\pi}{2}} e^{- 2\alpha^2 
R^2}
\sum_K \sum_J (-1)^{K+J} \kappa_K \kappa_J \nonumber \\ 
&&\hspace{-0.5cm} \times
\sum_{K1=0}^{min(K,J)} L^{(3/2)}_{K1}(0)
\big[ L^{(1/2)}_{K-K1}(2 \alpha^2 R^2) + L^{(1/2)}_{K-K1-1}(2 \alpha^2 
R^2)\big] \nonumber \\
&& \hspace{-0.5cm} \times
\bigg[ L^{(1/2)}_{J-K1}(2 \alpha^2 R^2)+ L^{(1/2)}_{J-K1-1}(2 \alpha^2 
R^2)\bigg].
\label{eq8} \end{eqnarray}

\noindent
where $\alpha = 1/b = \sqrt{m \omega/\hbar}$ is the inverse HO length, 
$K$ and $J$ are the principal 
HO quantum numbers of the shells, $\kappa_K$ and $\kappa_J$ the 
corresponding BCS amplitudes of the pairing tensor.

%\begin{equation}
%N(R)= \int dE \kappa (E) \int dp p?? |df_E(H_{cl.})/dp|??
%D(R)=\int dE \kappa (E) \int dp p?? |f_E(H_{cl.})|??
%\end{equation}

%\noindent
%where

%\begin{equation}
%\int dp p?? |df_E(H_{cl.})/dp|?? =....
%\int dp p?? |f_E(H_{cl.})|?? = ....
%\end{equation}

In the upper panel of Fig. 2 we show $\xi(R)$ for different values of $\eta$. 
It is seen 
that $\xi(R)$ only depends very weakly on the pairing strength for $\eta<1$, 
this  happens for instance around the 
minimum and the similarity with 
the results of the 
realistic calculations presented in \cite{pil07} and  displayed again in 
the lower panel 
of Fig. 2, is striking. 
In particular our model reproduces the very small value of $\xi(R)$ in the 
nuclear surface of about 2fm. For $\eta > 1$, the CL starts to move to 
lower values  in the interior. However, the minimum again only 
is very little affected.\\

\begin{figure}
\includegraphics[width=7.5cm,angle=-90]{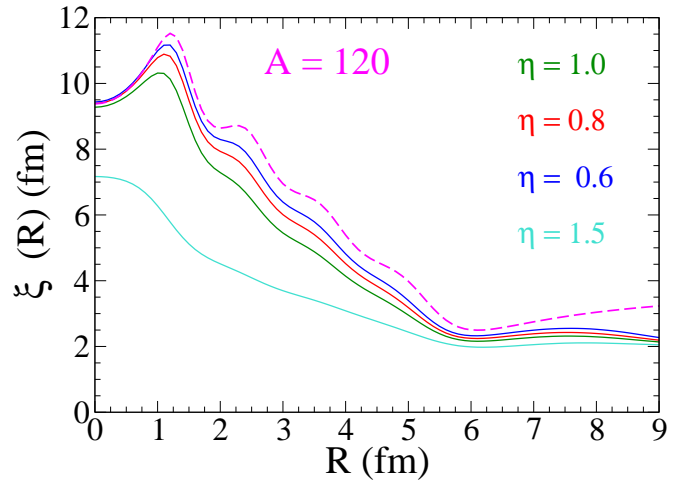}

\vspace{1.0cm}

\includegraphics[width=7.5cm,angle=-90]{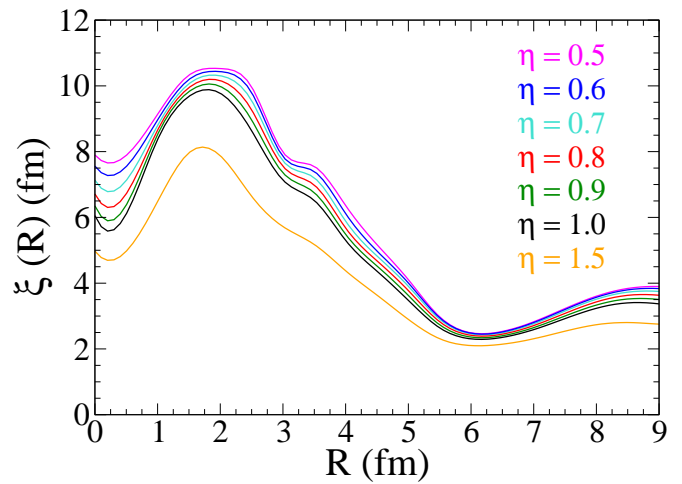}
\caption{(Color online)
Coherence length for different strengths of the pairing force as a 
function of the radial distance $R$ for a symmetric nucleus with A=120. The 
dashed line correspond to the no pairing limit (top). HFB coherence length
for the nucleus $^{120}$Sn computed for several strengths of the pairing 
D1S Gogny  force (bottom).} 
\label{Figure2}
\end{figure}

In our model it is now rather staightforward to understand where this striking 
{\it independence} of $\xi(R)$ on the intensity of pairing comes from. 
From Eq. (\ref{eq2}),
we can realize that the features of $f_{N_F}(H_{cl.})$, where $N_F$ 
corresponds to the major shell at the Fermi energy, have a width of 
order $\sim \hbar \omega$ (one may check this explicitly for some low 
order $L_N^{(2)}$ polynomials). 
Since in the case of nuclei $\Delta_N << \hbar \omega$, the $\kappa_N$ 
are essentially only active at the Fermi level and we 
approximately have from (3) 
that $\kappa({\bf R}, {\bf p})$ is proportional to $f_{N_F}(H_{cl.})$. In the 
limit $\eta \rightarrow 0$, we have the equality (strictly speaking one should 
in this limit change the name and not call it coherence length, 
since there is no coherence any longer; however, for convenience, we will 
not change the letter $\xi$ nor the name)

%\begin{eqnarray}
%&&\xi(R) = \sqrt{\frac{\int \frac{d^3p}{(2\pi \hbar)^3}
%[df_{N_F}(H_{cl.})/dp]^2}
%{\int \frac{d^3p}{(2\pi \hbar)^3}[f_{N_F}(H_{cl.})]^2}} 
%\equiv \sqrt{\frac{\mathcal {N}(R)}{\mathcal {D}(R)}}
%\nonumber \\
%&&=\sqrt{\frac{2\hbar^2}{m}}
%\sqrt{\frac{\int_{V(R)}^{\infty}
%d H_{cl.} \tilde{H}(H_{cl.},R)^3
%d H_{cl.}[H_{cl.} - V(R)]^{3/2}
%[df_{N_F}(H_{cl.})/dH_{cl.}]^2}{\int_{V(R)}^{\infty} dH_{cl.}
%[H_{cl.} - V(R)]^{1/2}
%\tilde{H}(H_{cl.},R)
%[f_{N_F}(H_{cl.})]^2}} \nonumber \\ 
%&&=\sqrt{\frac{2\hbar^2}{m}} \sqrt{\frac{\tilde{N}(R)}{\tilde{D}(R)}}
%\label{CL1}
%\end{eqnarray}
\begin{eqnarray}
\xi(R)&& {{\lim_{\eta \to 0}}\atop =} 
\sqrt{\frac{\int \frac{d^3p}{(2\pi \hbar)^3}
[df_{N_F}(H_{cl.})/d(p/\hbar)]^2}
{\int \frac{d^3p}{(2\pi \hbar)^3}[f_{N_F}(H_{cl.})]^2}}
\nonumber \\
&&= \frac{\hbar^2}{m}\sqrt{\frac{\int_{V(R)}^{\infty}
dH_{cl.} k_{H_{cl.}}^3(R)
[df_{N_F}(H_{cl.})/dH_{cl.}]^2}{\int_{V(R)}^{\infty} dH_{cl.}
k_{H_{cl.}}(R)
[f_{N_F}(H_{cl.})]^2}} \nonumber \\ 
&&= \sqrt{\frac{\int d^3ss^2\vert \rho_{N_F}({\bf R, s}) \vert ^2}{
\int d^3s\vert \rho_{N_F}({\bf R, s}) \vert ^2}},
\nonumber \\
\label{CL1}
\end{eqnarray}

\noindent
where $k_{H_{cl.}}(R)=\sqrt{\frac{2m}{\hbar^2}
(H_{cl.}-V(R))}$ and $\rho_{N_F}({\bf R}, {\bf  
s})$ 
is the Fourier 
transform of $f_{N_F}(H_{cl.})$ with 
respect to momentum $p$, that is the density matrix corresponding to the Fermi 
level $N=N_F$. The latter can be obtained from (2) as

\begin{eqnarray}
&&\rho_N({\bf R}, {\bf s}) = \frac{\alpha^3}{\pi^{3/2}} e^{-(R^2+\frac{s^2}{4})}
\times \nonumber \\
&&\sum_{K_1=0}^{K_1=N} (-1)^{N-K_1} L^{1/2}_{N-K_1}(2 \alpha^2R^2)
L^{1/2}_{K_1}(\frac{\alpha^2s^2}{2})
\label{DM} \end{eqnarray}

%where $k(H_{cl.},R)=[\frac{2m}{\hbar^2}(H_{cl.} - V(R))]^{1/2}$. 
\noindent
With a rescaling of the relative coordinate $s \to 2s$, we see the well 
known fact, see e.g. \cite{pil10}, that the density matrix for even/odd $N$ is 
completely symmetric/antisymmetric with respect to an interchange of relative 
and c.o.m. coordinates $s$ and $R$.
From Eq. (\ref{CL1}) it can be seen that the dependence 
of $\xi(R)$ on $\Delta$ has 
dropped out completely. This stems from 
the fact that in our H.O. model with its degenerate shells, in the 
limit $\eta \rightarrow0$, the chemical potential becomes locked 
exactly at the Fermi level, i.e. at the shell $N_F$. In general, this is not 
the case in a Woods-Saxon potential where it can happen that the chemical 
potential becomes situated in between two subshells. In the upper
panel of Fig. \ref{Figure2}, 
we 
also show the limiting value of the coherence length (broken line)
when $\Delta \to 0$. It is 
clear that this asymptotic form of the CL is very close to the other curves, 
in particular at the minimum. 
Therefore, in nuclear physics, {\it in what concerns the} CL, we are always 
almost in the asymptotic limit of vanishing pairing. 
In a sense, the closenenss of the CL to the corresponding uncorrelated value 
is one of the most striking manifestations that nuclei are in the weak 
coupling regime of pairing. Of course, this should not make us forget that 
on other quantities nuclear pairing has a strong influence. A particularly 
pertinent example, discussed recently \cite{mat05,pil10}, is the strong 
influence of 
parity 
mixing on the spatial features of the ({\it non-normalised}) pairing tensor.

Let us now try to analyse from where comes this typical behavior of the 
CL, i.e. of $\xi(R)$. It raises from $R$=0 up to $R$=1-2 fm, followed by a 
longer almost linear descent, 
passing through a shallow minimum of 2-2.5 fm, levelling off at some slightly 
higher asymptotic value. 
Before coming to this study, 
let us mention again that this behavior seems to be very robust 
being found in realistic HFB calculations in nuclei, see Fig. 3 in 
\cite{pil07}, in a slab 
geometry \cite{baldo}, as well as in the present very 
simplified HO model.

Let us consider the normalised square of the density matrix, 
as it enters the definition of the CL.  
From Eqs.(\ref{CL1}) and (\ref{DM})  we obtain 
\begin{eqnarray}
&&\frac{\vert \rho_N({\bf{R},\bf{s}}) \vert^2}
{\int d^3s  \vert \rho_N({\bf{R},\bf{s}}) \vert^2}
=\frac{\alpha^3}{4 \pi} \sqrt{\frac{2}{\pi}}
e^{-\frac{\alpha^2 s^2}{2}} \times \nonumber \\
&&\hspace{-1.0cm} \frac{\bigg(\sum_{K_1=0}^{K_1=N}
(-1)^{N-K_1} L^{(1/2)}_{N-K_1}(2 \alpha^2R^2)
L^{(1/2)}_{K_1}(\frac{\alpha^2 s^2}{2})\bigg)^2}
{\sum_{K_1=0}^{K_1=N} {\big(L^{(1/2)}_{N-K_1}(2 \alpha^2R^2)}\big)^2
L^{(1/2)}_{K_1}(0)}.
\label{eq6} \end{eqnarray} 
In the particularly simple case of $N$=1, $L^{(1/2)}_{1}(x)=\frac{3}{2}-x$
and $L^{(1/2)}_{0}(x)= 1$, and consqequently Eq.(\ref{eq6}), 
after multiplying by $s^4$, reads
\begin{eqnarray}
&&\frac{\vert \rho_1({\bf{R},\bf{s}}) \vert^2 s^4}
{\int ds \vert \rho_1({\bf{R},\bf{s}}) \vert^2 s^2}
=\alpha^3 \sqrt{\frac{2}{\pi}}
e^{-\frac{\alpha^2 s^2}{2}} \nonumber \\
&&\hspace{-1.0cm} \times 
\frac{ \alpha^4 (2R^2- s^2/2)^2 s^4}
{4 \alpha^4 R^4 - 6 \alpha^2 R^2 + 3.75}.
\label{eq7} \end{eqnarray} 
One sees that that (12) has an $R$ dependent node at $s=2R$, a feature 
which is important for interpreting the characteristic behavior of the CL.
After integrating (12) over s, one obtains for the CL 
\begin{eqnarray}
\xi^2(R)&&= \frac{\int ds (2R^2-s^2/2)^2 s^4e^{-\alpha^2 s^2/2}}{\int ds 
(2R^2-s^2/2) s^2 e^{-\alpha^2 s^2/2}}
\nonumber \\
&&=\frac{3}{\alpha^2}\times \frac{4x^2 - 20x + 35}{4x^2- 12x + 15}
\label{eq8a} 
\end{eqnarray} 
where $x = 2 \alpha^2 R^2$. Minimization with respect to $R$ implies that
\begin{equation}
4x^2 - 20x + 15 = 0,
\end{equation} \label{eq9}
\noindent
the roots of which are: $x_1 = 2.5 - \sqrt{2.5} = 0.9189$ and 
$x_2 = 2.5 + \sqrt{2.5} = 4.0818$, $x_1$ corresponding to the maximum 
and $x_2$ to the minimum.

 In the case of $N=1$, let us take, somewhat arbitrarily, the 
symmetric open 
shell nucleus $A=12$ (in order to fix the value of $\omega \propto  A^{-1/3}$). 
From the  definition of $x$ one 
obtains for
the position $R$ and the values of  maximum and  minimum of the CL 

\begin{equation}
R_{max} = 1.0315 fm \quad  \quad \xi(R)_{max} = 4.3476 fm,
\end{equation} \label{eq10}
and
\begin{equation}
R_{min} = 2.1740 fm \quad  \quad \xi(R)_{min} = 2.0629 fm.
\end{equation} \label{eq11}
%The coherence length corresponding to $A=12$ ($N=1$) is displayed in
%Fig.(3)
\begin{figure}
\includegraphics[width=7.5cm,angle=-90]{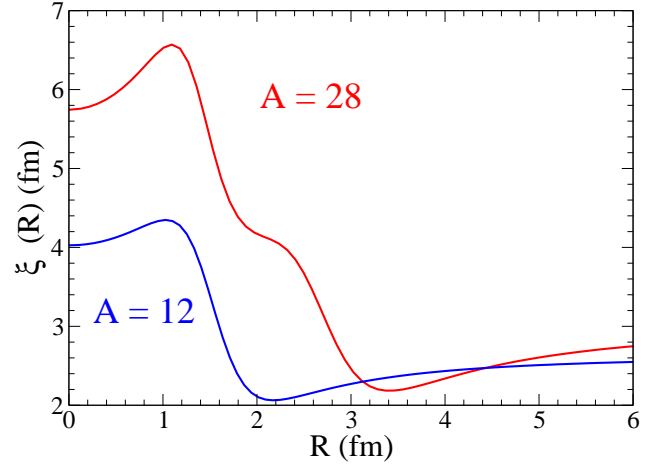}
\caption{Coherence length (fm) for systems containing A=12 and A=28
nucleons as a function of the distance to the center R (fm)} 
\label {CL12} \end{figure}

The coherence length corresponding to $A=12$ ($N=1$) is displayed in
Fig. \ref{CL12}.
It is very surprising that even for this simplest case of $N=1$ the essential 
features of 
the CL are already born out. For instance the minimal value is  
at about 2 fm, as in all other cases, realistic ones included \cite{pil10}. 
On the other hand for $N=0$, no $R$-dependence of the CL exists. The 
constant value of CL for $N$=0 is about 2 fm for, e.g. $A$=4, 
i.e. the $\alpha$ particle. Therefore, one needs at least to go to $P$-shell 
nuclei, i.e. $N=1$, so that in the interior the pairs can 
extend beyond  2fm. 
However, coming close to the edge of the nucleus, the rms. value of the 
pair gets (approximately) back to its value it has in the $\alpha$ particle.

The coherence length for $A=28$ what correspond to a mid shell nucleus with 
$N=2$ reads:
\begin{eqnarray}
&&\xi^2(R) = \frac{3}{\alpha^2} \times \nonumber \\
&&\frac{16x^4-224x^3+1160x^2-2312x+2009}{16x^4-160x^3+616x^2-888x+561},
\end{eqnarray} \label{eq12}
\noindent
where again $x = 2 \alpha^2 R^2$. The minimization with respect to $R$ 
implies that
\begin{eqnarray}
&&64x^6-1088x^5+7248x^4-27168x^3 \nonumber \\
&&+61340x^2-73348x+30435=0,
\label{eq13'}
\end{eqnarray} 
\noindent
which yields the only real roots: $x_1 = 0.7822$ and
$x_2 = 7.5622$, $x_1$ corresponding to the
maximum and $x_2$ to the minimum. The other four roots of (\ref{eq13'})
are complex.

From the previous definition of $x$ one obtains:
\begin{equation}
R_{max} = 1.0960 fm \quad  \quad \xi(R)_{max} = 6.5700 fm,
\end{equation} \label{eq13}
and
\begin{equation}
R_{min} = 3.4079 fm \quad  \quad  \xi(R)_{min} =  2.1838 fm.
\end{equation} \label{eq14}

\begin{figure}
\includegraphics[width=7.0cm,angle=-90]{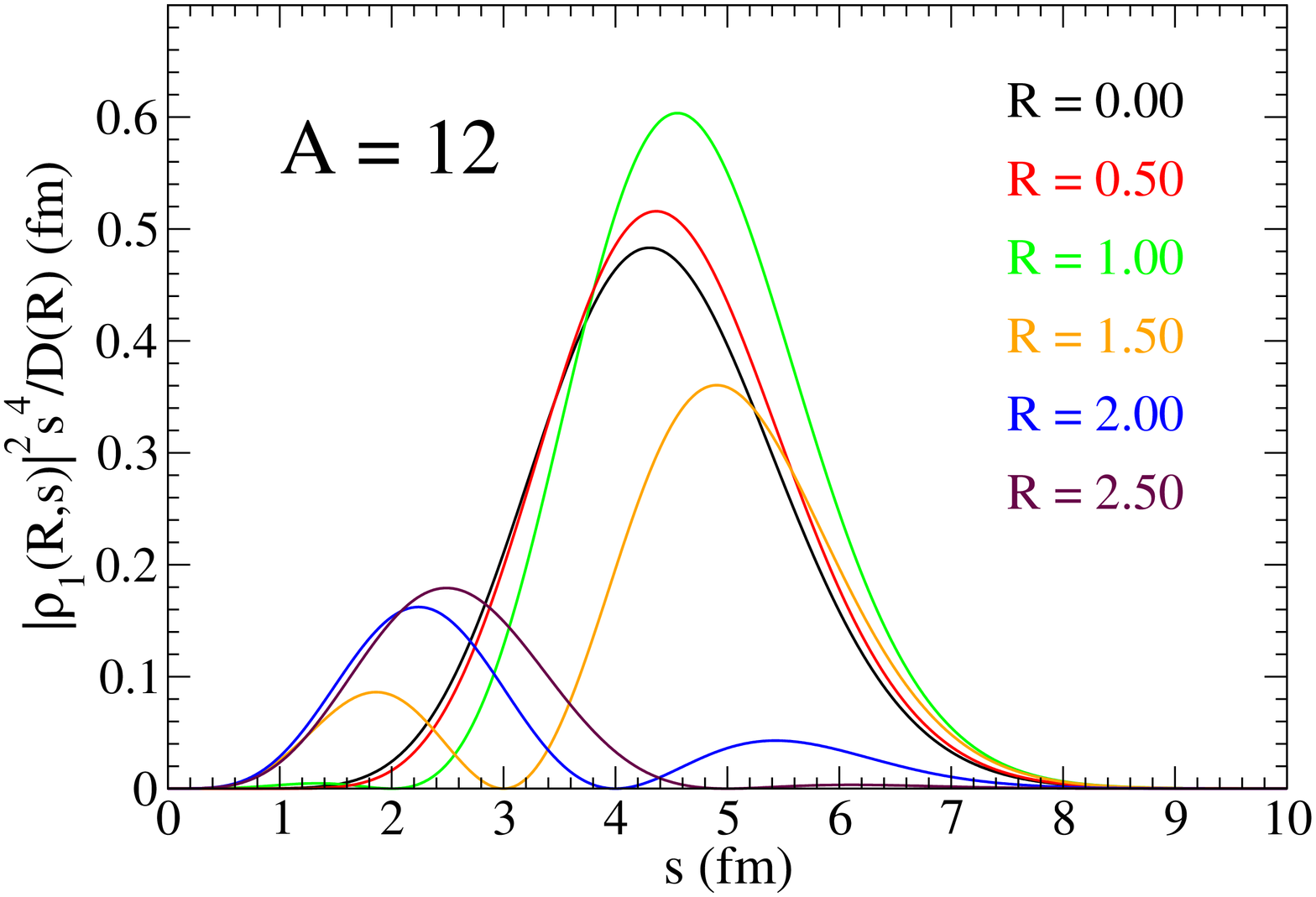}

\includegraphics[width=7.0cm,angle=-90]{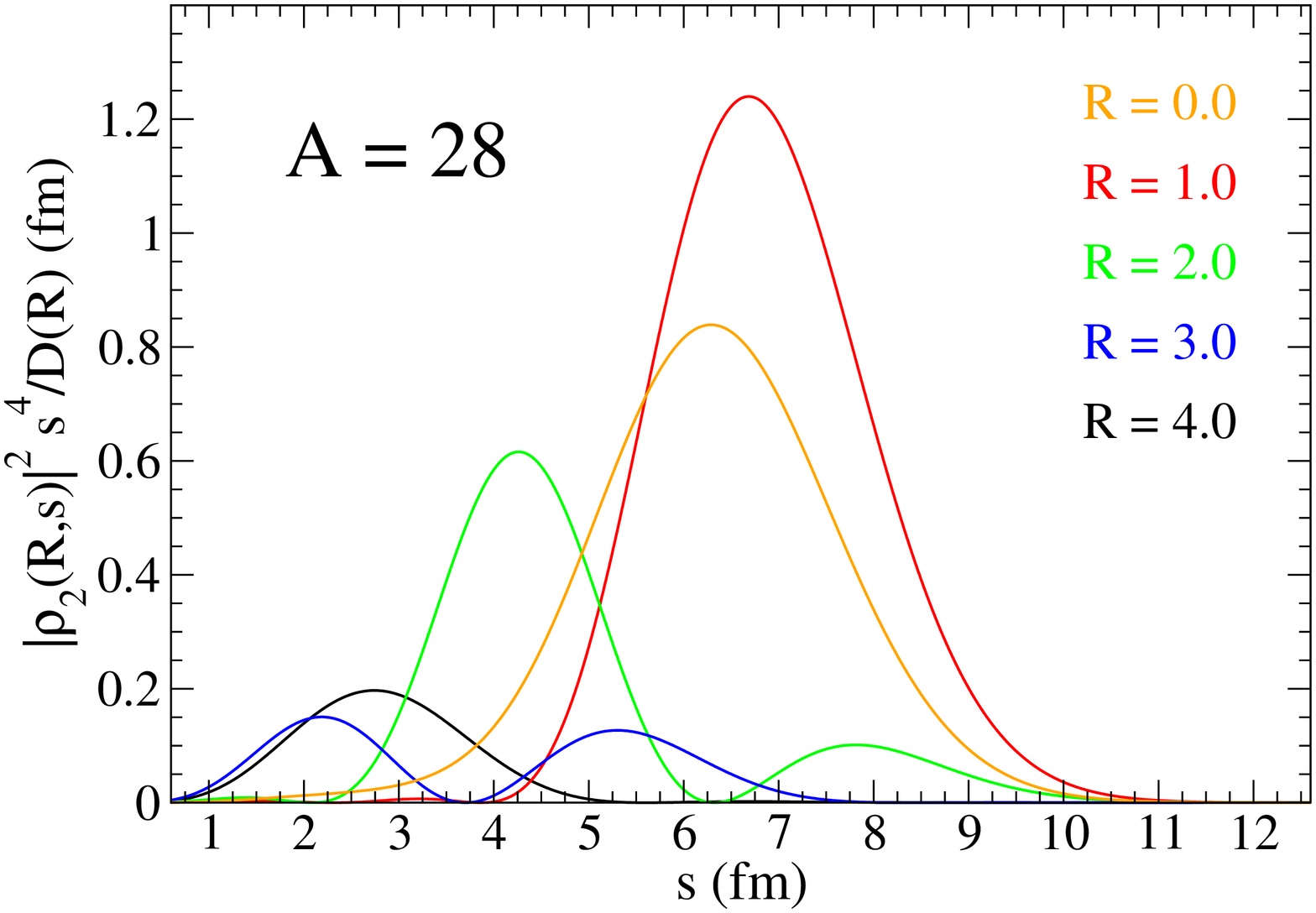}
\caption{(Color online) Normalized square of the density matrix 
$\rho({\bf{R},\bf{s}})$ multiplied by $s^4$ as a function of the relative 
coordinate $s$. In the upper panel it is displayed for A=12 ($N=1$)
and in the lower panel for A=28 ($N=2$)} 
\label{CL28} \end{figure}

The coherence length corresponding to this case is also displayed in
Fig.\ref{CL12}. In order to better understand the qualitatively similar 
behavior of the CL for the A=12 and A=28 cases, we show in Fig. \ref{CL28} 
expression (12) and the corresponding one for A=28 as a function of $s$ 
for various values of $R$.
The area below the curves in Fig. \ref{CL28} directly
yields the CL.
The striking feature is that the scenario is qualitatively 
much the same in both cases, inspite of the fact that for $N=2$ there 
are two nodes instead of one \cite{nodes}. The analysis shows that the two nodes also 
move proportional to $R$ from inside to outside in a similar way as for 
the $N=1$ case. We  surmise that the 
behavior stays more or less the same also for higher $N$ values. 
There are two asymptotic regimes where the nodal structure 
in $s$ practically does not influence the integrand in $s$ ( i.e. (12)) and 
which are more or less  dominated by a single bump structure. 
This is the case 
for very small $R$-values as well as for large $R$-values, approximately 
from the minimum point of the CL onwards. In between, the behavior switches 
from one regime to the other. This is where the CL shrinks about linearly 
with $R$. In order to exhibit the linear behavior more clearly, we show 
in Fig. 5 the CL's for A=12, 28, 120, and 8000. We scale in 
that figure the CL and the $R$ coordinate by the radius at the classical 
turning point, $R_t=\sqrt{\frac{2 \mu}{m \omega^2}}$, 
given by the intersection of the chemical potential $\mu = 46.933 MeV$ 
(remember that with a HO potential, $\mu$ is independent of of the 
nucleon number $A$) 
with the HO potential and, thus, representing the size of the system. 
It is seen that the different curves almost are 
superposed averaging around a linear descent. Only the beginning and the 
ends vary. The position of the 
minimum ranges between 
a little more than half of $R_t$ for A=12 to about 90 percent 
of $R_t$ for A=8000. In the interior, close to the origine, the pairs 
occupy the whole nuclear volume  while 
approaching the surface they steadily shrink to about 2-2.5 fm due to 
the close presence 
of the confinement. After the minimum, 
i.e. more or less after the classical turning point (for very small systems 
the latter does not have such a well defined meaning), the pair wave 
function enters 
the evanescent region and again slightly expands before reaching the 
asymptotic 
value. It also is worth mentioning that the minimum of the CL is very 
slowly increasing 
with particle number, approximately as $\sim A^{1/6}$. 
For $A=8000$ the minimum value is 
about 4fm. Once $\omega \rightarrow 0$, the CL approaches infinity everywhere. 
A generic feature also is that, independent of $A$, starting from the 
center at $R=0$, 
the pairs first slightly expand up to $R \sim$ 1fm before becoming smaller 
getting closer to 
the border of the mean field. For these $R$-values around the origine, 
the nodes lie in  the region which is 
dominated by the phase space factors $s^2$ and $s^4$ in the integrals 
over $s$ in (13), i.e. well to the left of the maximum of the bump 
created by the function $s^4 e^{-\alpha^2 s^2/2}$. 
It can easily be verified from our 'easy' example $N=1$, eqs (12) 
and (13), that for very small values of $R$, the surface corresponding to 
the $s$-integral of the 
denominator decreases faster than the one of the numerator. Therefore, 
the CL increases. However, once the node comes into the region where the 
exponential regime takes over, i.e. where the extension of the system is felt, 
the CL starts its regression. These 
considerations may be elaborated in all details for the case $N=1$ and also 
further be elucidated in considering as a complement 
to the density matrix $\rho_N({\bf R}, {\bf s})$ its Wigner representation 
Eq (\ref{eq2}). Not to make the present discussion too heavy, we refrain 
from entering 
these more detailed considerations. The case $N=1$ is, as seen, already 
characteristic and can be studied straightforwardly. 

\begin{figure}
\includegraphics[width=7.5cm,angle=-90]{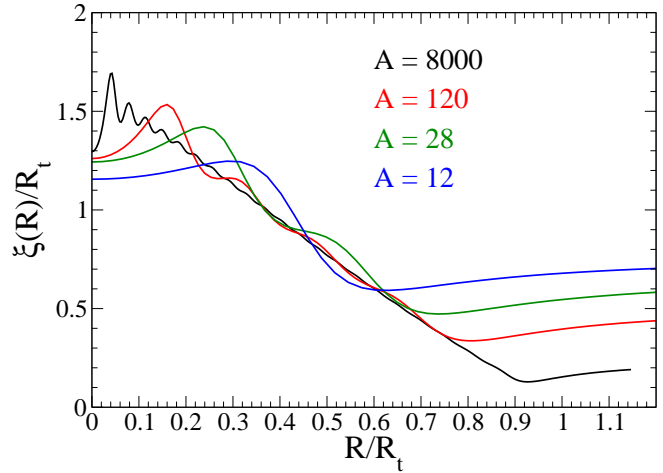}
\caption{(Color online)
Coherence length (CL) for systems containing $A$= 12, 28, 120 and 8000
nucleons as a function of the radial distance $R$. Notice that 
both, coherence length and radial distance, have been scaled by the 
classical turning 
point distance $R_t$ (see text)} 
\label{Figure6}
\end{figure}
We also should mention that even in our averaging over major shells orbit 
mixing {\it within} the shell takes place.  The cross terms give raise to a 
distructive interference still lowering the minimum of the CL by 
a small but definite amount of about 0.5 fm from its non 
averaged values. This can be realised in comparing Fig. 6 where the CL, i.e.
local in $R$ rms radii 
from individual HO orbits are displayed (for a precise definition, see 
\cite{pil10}) 
with the broken line in the upper panel 
of Fig. 2. 
%{\bf Comment: No rms radii of individual orbits is displayed in Fig. 6!.
%What is displayed is the contribution of each orbit to the CL. In the 
%upper panel of Fig. 2 again what is displayed is the total CL. at least 
%the phrase like it is, is confuse.} 

Intrashell averaging, therefore, 
is present even in the limit of very small pairing with gap values of the 
order of subshell spacings. In \cite{pil10} the same study is performed with 
the self consistent HFB orbits, see Fig. 17 in that reference. It is seen 
that in the self consistent case the reduction of the minimal value of the CL 
from intra shell mixing is 
about 30 percent and, therefore, somewhat stronger than in the present 
simplified HO model. 
Though not completely negligeable, 
this interesting behavior is nontheless a minor effect with respect to the 
feature we are discussing in this work, namely a surprising reduction of 
the minimal 
value of the CL by a factor 3-4 from a simple weak coupling 
estimate \cite{BM75}.

\begin{figure}
\includegraphics[width=6.5cm]{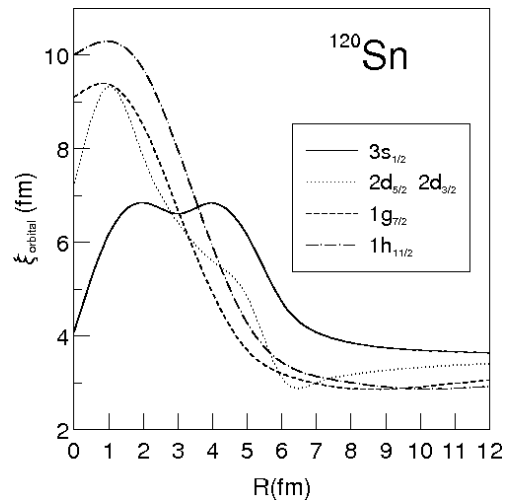}
\caption{
Individual contributions to the CL in the HO model for A=120 corresponding to 
the different individual orbitals of the last shell.}
\label{Figure8}
\end{figure}

We, therefore, can say that in cases where in finite Fermi systems typical 
values of gap parameters around the Fermi energy are smaller 
than $\hbar \omega$ or energy differences between neighboring 
major shells in spherical nuclei, the size of Cooper pairs in 
superfluid nuclei, or other finite Fermi systems, is essentially 
determined by the spatial extension of the 
single particle wave functions close to the Fermi energy. Passing for 
the sake of the argument to a continuum version of (2), 
i.e. $\kappa({\bf R}, {\bf p})=\int dE \kappa(E) f_E(H_{cl.})$, we 
see that in nuclei where the width of $\kappa(E)$ is much smaller than 
the width of $f_E(H_{cl.})$, the CL is dominated by $f_E(H_{cl.})$ on 
the Fermi surface.  Of course, a very 
different situation prevails in the opposite regime 
where $\Delta >> \hbar \omega$. In the extreme case of infinite matter 
or $\hbar \omega \rightarrow 0$,  we 
have $f_E(H_{cl.}) \rightarrow \delta(E-H_{cl.})$, and the ratio
of the values of the widths is inversed! Simple 
scaling arguments show 
that in the latter case $\xi \sim 1/\Delta$ which  also is reflected in 
the well known expression given by Pippard \cite{FW71}

\begin{equation}
\xi = \frac{1}{\pi}\frac{ \hbar^2}{m}\frac{k_F}{\Delta}
\end{equation}

\noindent
or by an equivalent formula given in Appendix of our preceding paper 
\cite{pil10}.
Therefore, in the infinite matter case the dependence on the gap is not at 
all compensated between numerator and denominator in eq (6), 
whereas this 
is the case in finite nuclei, see eq (9).

\noindent
%Of course, using the local density approximation, is equivalent to  
%the infinite matter regime 
%showing that  LDA is not valid to estimate the coherence length 
%in finite nuclei. 
As a consequence, the use of LDA, which 
is equivalent to the infinite matter regime, is not valid to estimate the 
coherence length in finite nuclei. 
For other quantities, however, as, e.g. the pairing energy, 
LDA gives a reasonable good average \cite{kur89}. Nevertheless, even in such 
favorable cases, LDA is very much  at the limit of  its validity, for instance 
in what concerns a detailed description of the radius dependence of various 
pairing quantities. Further considerations on this subject will be published 
elsewhere \cite{vinas-farine}.\\
In conclusion, concerning the extension of Cooper pairs in finite 
superfluid Fermi systems, we have identified two regimes: 
One for $\hbar \omega >> \Delta$ where the coherence length is
practically 
independent of $\Delta$ and determined by the spatial extension of the 
single particle wave functions. Besides in nuclei, such a situation may be 
found in ultrasmall superconducting metallic grains \cite{farine,delft}. 
In the second regime 
with $ \hbar \omega << \Delta$, the coherence length is approximately 
inversely proportional to the gap values. The latter situation is, besides 
nuclear matter, e.g. 
realised in cold superfluid fermionic atoms in traps where typical values 
of $\Delta/(\hbar \omega)$ may be of the order of ten or even 
larger  \cite{string}. 
 It would be interesting to study the cross over from one regime to the 
other in more detail.

Let us finally wrap up the situation of the CL in nuclei. We found that a 
simplest spherical HO model already simulates quite faithfully realistic 
HFB calculations with the Gogny force. In what concerns the CL, the situation 
for nuclei is such that there is very little difference between rms. values 
of uncorrelated pairs coupled to $L=S=0$ calculated locally as a function of 
the radius $R$ and local Cooper pair sizes calculated with the nominal pairing 
interaction. Therefore, the small Cooper pair size of 2.0-2.5 fm in the 
surface of nuclei is practically entirely a finite size effect and has not 
much to do with existing enhancement of pairing in the nuclear surface. A very 
characteristic and generic pattern has emerged. In the lightest nuclei, 
like, e.g.,
the $\alpha$ particle, their size is so small that the extension of a pair 
cannot reach more than 2 fm. Going to $P$ shell nuclei, in the 
interior the pairs can already somewhat extend but approaching the border 
of the mean field 
they shrink until they again reach a value of around 2 fm due to the resricted 
space around the surface. In the interior the pairs grow approximately with the 
size of the nucleus, see Fig. 5 but towards the surface they always regress to 
their very small value. In the  evanescent region, 
the pair sizes 
become slightly larger than their minimum value in the surface region but this 
increase is very moderate. A characteristic feature also is that the pairs only 
feel the finite size from $R$=1 fm onwards. Before, they slightly expand up 
to $R$=1 fm, independent of the mass number of nuclei and of parity 
of the shell. This scenario of a 
first slight increase, followed by a longer linear descent, 
before going through a 
shallow minimum at 2.0 - 2.5 fm, levelling off in a slightly increased 
asymptotic value is practically a generic feature of local pair sizes in 
nuclei. It is seen in our schematic model, but also in realistic calculations, 
see Fig. 3 of \cite{pil07}, though in the latter case some scatter exists, 
probably due to more pronounced shell effects. This characteristic pattern of 
local Cooper pair sizes, practically independent of the strength of the pairing 
force as long as it stays below the nominal value, is one of the clearest 
theoretical manifestations that nuclei are in a weak coupling regime 
characterised by gap values $\Delta << \hbar \omega$. In the opposite 
limit $\Delta >> \hbar \omega$, as prevails in infinite matter, but in the 
regime where $\Delta << \mu$ that is still in weak coupling 
\cite{weakcoup}, the coherence 
length varies inversely proportional to the gap, a fact which is well known. 
The fact that Cooper pair sizes are largely dominated by geometry should, 
however, not make us forget that for other quantities 
nuclear superfluidity has an enormous impact. To say it again, a 
particularly striking 
example, besides others more standard ones, is the effect of parity mixing on 
the spatial behavior of the non local (unnormalised) pairing tensor, as 
revealed recently \cite{mat05,pil10}.
Indeed, this not normalised pairing tensor $\kappa(\vec{R},\vec{s})$ 
becomes very much localised in $\vec{s}$ around the $\vec{R}$-axis 
whereas 
the parity projected $\kappa(\vec{R},\vec{s})$ is completely 
delocalised \cite{mat05,pil10}
However, due to normalisation in the coherence length this feature 
is cancelled out.

\begin{acknowledgments}
We appreciated very stimulating discussions with J.-F. Berger, A. Pastore, 
and N. Sandulescu. 

Work partially supported by the IN2P3-MICINN 
agreement FPA2008-03865-E/IN2P3
and by the Spanish Consolider-Ingenio 2010
program CPAN CSD2007-00042.
X.V. also acknowledges the support from FIS2008-01661 
(Spain and FEDER) and 2009SGR-1289 from
Generalitat de Catalunya (Spain). 
\end{acknowledgments}

%\begin{figure}
%\includegraphics[width=12cm,angle=-90]{FHO120f.ps}
%\caption{\label{Figure1}}
%\end{figure}

%\begin{figure}
%\includegraphics[width=12cm,angle=-90]{KappaRS4.ps}
%\caption{\label{Figure2}}
%\end{figure}

%\begin{figure}
%\includegraphics[width=12cm,angle=-90]{archsnew1.ps}
%\caption{\label{Figure1}}
%\end{figure}

%\begin{figure}
%\includegraphics[width=12cm,angle=-90]{CLHO.ps}
%\caption{\label{Figure3}}
%\end{figure}

%\begin{figure}
%\includegraphics[width=12cm,angle=-90]{FHO120e.ps}
%\caption{\label{Figure4}} 
%\end{figure}

%\begin{figure}
%\includegraphics[width=12cm,angle=-90]{RMSP2.ps}
%\caption{\label{Figure4}}
%\end{figure}

%\begin{figure}
%\includegraphics[width=12cm,angle=-90]{CLDELTA.ps}
%\caption{\label{Figure5}}
%\end{figure}


\begin{thebibliography}{2007}
\bibitem{nodes}
As a matter of fact for values of $R < \frac{1}{2 \alpha}$,
no real nodes exist. This, however, does not perturb the general pattern
of lower panel of Fig. 4 being qualitatively similar to the one of the 
upper panel.
\bibitem{weakcoup}
The criterion $\Delta/\mu \sim \Delta/\varepsilon_F << 1$ is, in the 
continuum limit, equivalent to the usual weak coupling estimate $G \rho$ 
(matrix element times level density at Fermi energy) $<< 1$. However, in 
finite systems with shell structures, the additionla distinction 
$\Delta/(\hbar \omega) \gtrless 1$ can be made influencing the details of 
pairing. A nice discussion of weak vs strong pairing in finite systems is 
given in \cite{delft}.
\end{thebibliography}
\end{document}